\documentstyle[twocolumn,aps]{revtex}
\begin{document}
\draft
\title{\bf {Violation of general Friedel sum rule in mesoscopic systems}}
\author{P. Singha Deo\cite{eml}} 
\address{S. N. Bose National Centre for Basic Sciences, JD Block, Sector 3,
Salt Lake City, Calcutta 91, India.}
\maketitle
\begin{abstract}
In the wake of a new kind of phase generally occurring in mesoscopic
transport phenomena, we discuss the validity of Friedel sum rule
in the presence of this phase. We find that the general Friedel
sum rule may be violated.
\end{abstract}
\narrowtext

With large scale research in Mesoscopic Physics over the last
few decades,
many of the well established notions of Condensed Matter Physics
has been found to be violated in mesoscopic samples. Breakdown
of Onsager reciprocity relation \cite{but}, violation of 
Ohms law \cite{kum}, absence of 
material specific quantities like resistivity \cite{sto}, violation of
Hund's rule \cite{kos} etc., are a
few such examples. The purpose of this work is to show
the violation of Friedel sum rule in mesoscopic systems.

Friedel sum rule relates the density of states inside a
fixed potential scatterer
to the scattering phase shifts \cite{heg}. 
A deduction of the sum rule
can be found in many text books \cite{boo1,boo2} and intuitively
it can be understood as follows. Consider for example a fixed
spherically symmetric potential scatterer. Now we enclose
it in a larger spherical volume. In an energy interval $dE$,
the number of states depend on the number of times the
specific boundary conditions can be fulfilled by the
wave function of the electron. So when the energy is changed,
it introduces a phase shift of the electron wave function
and so changes the number of times the specific boundary condition
can be satisfied. And hence the density of states $\rho '$ inside
the impurity is related to the scattering phase shift $\eta$ in
the following fashion \cite{boo1}.

\begin{equation}
{\partial \eta \over \partial E}=\pi \rho '.
\end{equation}
This can be extended to the partial wave analysis of scattering states
and many important issues can be understood in terms of the Friedel
sum rule \cite{boo2}. In case of a non-spherical scatterer or non-spherical
Fermi surface,
the scattering matrix is in general an NxN matrix. For any
such general NxN scattering matrix $S$, the Friedel sum rule can be
written as \cite{lan}

\begin{equation}
{\partial \theta/\partial E}=\pi \rho ',
\end{equation}
where $\theta=\Sigma_i^N \xi_i$, $exp[2i\xi_i]$ being the eigenvalues
of the scattering matrix $S$. This can be further written in a compact form
as

\begin{equation}
{1 \over 2i}{\partial \over \partial E}(ln(det[S]))= \pi \rho '.
\end{equation}
For one-dimensional systems
where the scattering matrix is 2x2, the Friedel sum rule was thought to
be further simplified to give \cite{har}

\begin{equation}
{\partial arg(t) \over \partial E}=\pi \rho ',
\end{equation}
where $t$ is the transmission amplitude but this is not true.

Recently a new phase has been discussed in Ref. \cite{deo1} for
scattering by a stub where the scattering matrix is 2x2, 
and it is believed \cite{the,deo2,lee,tan} that this phase is
also observed in mesoscopic systems experimentally \cite{sch}. 
This phase is a general feature of transmission
zeroes that always occur in Fano resonances in
Quantum Wires and Dots, the stub structure being
the simplest example \cite{deo2,lee}.
It was shown in Ref. \cite{deo1}
that the phase slips is a new phase associated with the
violation of parity effect because it is different
from Aharonov-Bohm phase, statistical phase and phase due to
wave-like motion of electrons depending on their 
wave vector or energy.
Had it not been different from the other three phases,
parity effect would not have been violated \cite{deo1}.
The specialty of this phase
is that it is discontinuous as a function of energy, i.e.,
the phase of the wavefunction changes by $\pi$ although its energy
does not change. 
To be more precise this phase does not originate from change in
wave-vector due to change in energy. Hence in view of the
discussions before Eq. 1 one can question the validity
of Friedel sum rule in the presence of this phase \cite{deo2,lee,tan}.
We shall give a pictorial description of this phase later 
(short-dashed and long-dashed curves in Fig.~1).

The scattering matrix for the stub is

\begin{equation}
S=\pmatrix {{r} & {t} \cr
{t} & {r} \cr}
\end{equation}
where $r$ and $t$ are reflection and transmission amplitudes across
the stub and are
\begin{equation}
r=cos[kL]/(-cos[kL] + 2i sin[kL])
\end{equation}
and
\begin{equation}
t=(-2i sin[kL])/(cos[kL] - 2i sin[kL]).
\end{equation}
The eigenvalues of the S matrix are
\begin{equation}
(cos[kL] + 2i sin[kL])/(-cos[kL] + 2i sin[kL]) \quad and \quad -1
\end{equation}
Hence as defined in Eq. 2
\begin{equation}
\theta={1 \over 2}
ArcTan[-4 cos[kL]sin[kL]/(-cos[kL]^2 + 4 sin[kL]^2)]
\end{equation}

In Fig.~1 we plot $\theta$ (solid curve), $arg(t)$ 
(short-dashed curve) and $arg(r)$ (long-dashed curve)
(given in Eqs. 6, 7 and 9) versus $kL$.
It can be seen that $arg(t)$ and $arg(r)$ show discontinuous
jumps and drops by $\pi$ \cite{deo1}
but they cancel in such a way
that $\theta$ is continuous and monotonously increasing. 
Hence one finds that the Friedel sum rule (Eqs. 2 and 3) is
not violated \cite{tan} although because of the discontinuous
slips in $arg(t)$ Eq. 4 is obviously violated
because density of states can never be
infinite while the LHS of Eq. 4 can be infinite. 
And hence one can say that
so far no one has found a violation of
Friedel sum rule (Eqs. 2 and 3).
We shall show the violation of the
Friedel sum rule in the presence of this new phase.

Transport across the stub structure has acquired a lot of importance
recently \cite{the,lee,tan,bay}. 
All analysis so far are
based on calculations with a hard wall boundary condition
(an infinite step barrier potential or an infinite step well potential)
at the dead end of the stub (we refer to it as the hard walled
stub and for which Eqs. 6, 7, 8 and 9 are
derived). An infinite potential well at the dead end of the stub
reflects an incident electron with unit probability.
Now a small perturbation from this would be a finite but very
deep potential well at the dead end of the stub (soft walled stub).
Electrons are almost entirely reflected from the
end of the stub and a negligible fraction escapes. 
Dephasing can also give similar escape probability.
The scattering problem in this case is depicted in Fig.~2 and
also explained in the figure caption. It is solved
using the mode matching technique or Griffith's boundary
conditions \cite{jay1}, that give the continuity of wavefunction
and the conservation of currents at the junctions.
In this case the transmission zero in x-direction
is replaced by a minimum \cite{jay2}.
We first intend to understand
what happens to the discontinuous phase change that
occur due to transmission zeroes in this case.
So in Fig.~3 we plot transmission coefficient
$T=|t|^2$ (solid curve) and the argument of the transmission
amplitude $t$ (short-dashed curve) in x
direction, versus $kL$ for an almost hard walled stub.
The transmission coefficient shows very deep minima
and at the same points $arg(t)$ show 
very sharp but continuous drops.
For the completely hard walled stub there is an
exact zero and associated with it a
discontinuous slip by $\pi$ as shown in Fig.~1.
In the same figure (Fig.~3) 
we also plot transmission coefficient
in the x-direction
(dash-dotted curve) and the corresponding
argument of the transmission amplitude
(long-dashed curve) versus $kL$ for a very soft walled stub.
At the points where the solid curve show very deep minima,
dash-dotted curve show shallow minima. Also 
the fast phase drops
change over to a slower decrease.

Having understood the phase slips further we move on to
the three prong scatterer (Fig.~4) that is often encountered
in mesoscopic systems \cite{gan}
including the experimental set up of Ref. \cite{sch}
and many such similar experiments. 
The scattering problem in this case is described in
the figure caption.
From the continuity of wavefunctions (first Griffith's boundary
condition) we get the following equations (variables
and parameters are defined in Fig.~4 and it's caption).
$$1+r=a\,exp[-iqL_1]+b\,exp[iqL_1]; \quad a+b=c+d; $$
$$a+b=f+g, \quad c\,exp[iqL_2]+d\,exp[-iqL_2]=e;$$
\begin{equation} 
f\,exp[iqL_3]+g\,exp[-iqL_3]=h.
\end{equation}
And from the second Griffith's 
boundary condition which is the conservation
of currents at the junctions ($\Sigma_i {d \psi_i \over dx_i}=0$,
that can be derived from current conservation, here $\psi_i$
is a wavefunction at a junction, $x_i$ is coordinate at that
junction, and the sum over $i$ stands for all such wavefunctions
incoming or outgoing at a junction, the convention followed is that
currents flowing into the junction is positive while
currents flowing out of the junction is negative) 
we get the following equations.
$$k-kr-qa\,exp[-iqL_1]+qb\,exp[iqL_1]=0;$$
$$a-b-c+d-f+g=0;$$
$$qc\,exp[iqL_2]-qd\,exp[-iqL_2]-ke=0;$$
\begin{equation}
qf\,exp[iqL_3]-qg\,exp[-iqL_3]-kh=0.
\end{equation}
Thus we have 9 equations and exactly 9 unknown
quantities ($a,b,c,d,e,f,g,h$ and $r$) and so
the problem is completely defined. Once the unknowns are solved,
the wavefunction is known at all points exactly and so
the density of states as well as the scattering matrix
can be calculated exactly.
The scattering matrix in this case is
\begin{equation}
S=\pmatrix {{t_{11}} & {t_{12}} & {t_{13}} \cr
{t_{21}} & {t_{22}} & {t_{23}} \cr
{t_{31}} & {t_{32}} & {t_{33}} \cr}.
\end{equation}
Here $t_{11}$=$r$=transmission amplitude to the first
prong when the incident beam is from the first prong.
$t_{12}=e$ is the transmission amplitude to the second
prong when the incident beam is from the first prong.
$t_{13}=h$ is the transmission amplitude to the third
prong when the incident beam is from the first prong.
The other matrix elements are to be calculated when
similar incident beam of unit flux
is from the other two directions in Fig.~4.
For the case of Fig.~4, the partial density of states 
is given by the following expression
$$\rho_1=\pi \rho_1 ' ={\pi \over hv} [
\int_{-L_1}^0|aexp[iqx]+bexp[-iqx]|^2 dx +$$
$$\int_{0}^{L_2}|cexp[iqy]+dexp[-iqy]|^2 dy +$$
\begin{equation}
\int_{0}^{L_3}|fexp[iqz]+gexp[-iqz]|^2 dz ],
\end{equation}
where $v=\hbar k/m$. $a,b,c,d,f$ and $g$ are determined
from Eqs. 10 and 11.
$\rho_2$ and $\rho_3$ are to be evaluated similarly
when the incident beam is from the other two directions
in Fig.~4, and $\rho=\rho_1+\rho_2+\rho_3$.
For the symmetric three prong scatterer ($L_1=L_2=L_3=L$),
the antiresonances
are almost cancelled by the resonances but still a violation of 
Eq. 2 or 3 can be seen at low energy.
This cancelling effect of resonance and
antiresonance can be avoided by choosing incommensurate
values of ($L_1+L_3$) and $L_2$, i.e., for asymmetric
configurations.
We will now go to the asymmetric configuration and demonstrate
a large difference between $\tau=\partial \theta /\partial E
={1 \over 2i}{\partial \over \partial E}[ln[Det[S]]$
and $\rho=\pi \rho'$ at large energies ($E \approx V$). 
This is shown in Fig.~5. We want to emphasize that
at very high energy, compared to the
energy scale $V$ in the system, 
when multiple scattering and the new phase
becomes insignificant, we recover Friedel sum rule perfectly.
But when this new phase is present at energies ($E<V$), there is
a large difference between $\tau$ and $\rho$ and hence
a complete violation of Friedel sum rule.
In Fig.~5, $\tau$ or $\partial \theta /\partial E$,
can become substantially negative, i.e., $\theta$
can undergo a drop like $arg(t)$ in Fig.~3. 
The new phase need not always appear as a drop
but can also appear as a rise and then the LHS of
Eq. 3 can remain positive all the time while deviating
from the RHS of Eq. 3. This is shown in Fig 6.

Thus our exact calculation of density of states and scattering matrix
elements show the deviation of ${1 \over 2i}
{\partial \over \partial E}ln[Det[s]]$ from $\pi \rho'$ in the
presence of phase slips. The phase slips are a general feature
of Quantum wires with defects \cite{deo2} and Quantum Dots and these
phase slips are at the origin of drops in $\theta$
and hence deviation
or violation of Friedel sum rule.
Only 2x2 S-matrix is a special case where as shown in Fig. 1
some scattering matrix elements undergo a phase jump
and some undergo a phase drop in such a manner that
they cancel and the phase slips
do not affect $ln[Det[S]]$ or $\theta$. The general feature is
that they do not cancel and Friedel sum
rule gets violated. The attractive potential in
the three prong scatterer offsets the
symmetry between the phase jumps and the phase drops
so that they do not cancel each other.

The author acknowledges useful discussions with Prof. M. Manninen.

\centerline{Figure captions}

\noindent {\bf Fig.~1} Arg(r) (long-dashed curve), arg(t) (short-dashed
curve) and $\theta$ (solid curve) for the hard walled
stub. Length of the stub is $L$ and it is taken to be the
unit of length. We choose $\hbar=2m=1$.

\noindent {\bf Fig.~2} A scattering problem with conventional notations
is depicted here. $k=\sqrt E$ is the wave vector in the thin regions
where the Quantum Mechanical potential is 0. $q=\sqrt{E+V}$ is
the wave vector in the thick regions where the Quantum Mechanical
potential is $-V$. x and y are coordinates and the origin
of coordinates is also depicted in the figure. t and c are
transmission amplitudes in x and y directions, respectively, while
r is the reflection amplitude. Distance between points
P and Q is L.

\noindent {\bf Fig.~3} The solid curve is transmission coefficient T
=$|t|^2$ across the soft walled stub described in Fig.~2.
The short-dashed curve is the phase of the transmission amplitude
$t$ across the stub. We choose $VL^2=-10^6$ so that it is
in the hard wall limit, and $\hbar=2m=1$.
Next we make $VL^2$=-100 and plot the transmission coefficient T
in dash-dotted curve. The phase of the transmission amplitude t
is given by long-dashed curve.

\noindent {\bf Fig.~4} A scattering problem with conventional notations
is depicted here. $k=\sqrt E$ is the wave vector in the thin regions
where the Quantum Mechanical potential is 0. $q=\sqrt{E+V}$ is
the wave vector in the thick regions where the Quantum Mechanical
potential is $-V$. x,y,z,u,v and w are coordinates and the origin
of coordinates is also depicted in the figure. e and h are
transmission amplitudes in respective directions, while
r is the reflection amplitude. Distance between (u=0) and
(x=0,y=0,z=0) is $L_1$. 
Distance between (v=0) and
(x=0,y=0,z=0) is $L_2$.
Distance between (w=0) and
(x=0,y=0,z=0) is $L_3$.

\noindent {\bf Fig.~5} $\rho$ (solid curve) and $\tau={d\theta \over dE}$
=LHS of Eq. 3
(dotted curve) versus $kL$ for the scattering problem
described in Fig.~4. We choose $VL^2=-100$, $L_1=L_3=L$, $L_2$=4$L$ 
and $\hbar=2m=1$.

\noindent {\bf Fig.~6} $\rho$ (solid curve) and $\tau={d\theta \over dE}$
=LHS of Eq. 3
(dotted curve) versus $kL$ for the scattering problem
described in Fig.~4. We choose $VL^2=-100$, $L_1=L_3=L$, $L_2$=2.4$L$ 
and $\hbar=2m=1$.

\begin{thebibliography}{99}
\bibitem[*]{eml} Electronic mail: deo@boson.bose.res.in
\bibitem{but} M. B{\"u}ttiker, Phys. Rev. Lett. {\bf 57}, 1761 (1986).
\bibitem{kum} N Kumar and A M Jayannavar Phys. Rev. B 
{\bf 32}, 3345 (1985); 
B. J. van Wees et al, Phys. Rev. Lett. {\bf 60}, 848 (1988);
D. A. Wharam et al, J. Phys. C {\bf 21}, L209 (1988).
\bibitem{sto} A. D. Stone and A. Szaffer, IBM Journ. of Res. Development,
{\bf 32}, 384 (1988).
\bibitem{kos} M. Koskinen, M. Manninen and S. M. Reimann, 
Phys. Rev. Lett. {\bf 79} 1817 (1997).
\bibitem{heg} For charge neutral systems there is a one to
one correspondence between density of states and accumulated charge
in the field of a scatterer. In that case Friedel sum rule
also relates scattering phase shifts to the displaced charge
(see Ref. \cite{boo1}).
There can be situations when charge neutrality is violated
as in the case of R. Egger and H. Grabert,
Phys. Rev. Lett. {\bf 79}, 3463 (1997). However we will show a violation
of Friedel sum rule at a more fundamental level i.e.,
density of states itself starts deviating from that predicted
by Friedel sum rule.
\bibitem{boo1} J. M. Ziman, Principles of Solids, 2nd ed.,
Cambridge University Press, 1972.
\bibitem{boo2} W. Jones and N. H. March, Theoretical Solid State
Physics, Vol. 2, Dover Publications, Inc, New York, 1973.
\bibitem{lan} J. S. Langer and V. Ambegaokar, Phys. Rev. {\bf 121},
1090 (1961).
\bibitem{har} W. A. Harrison, in {\it Solid State Theory} (Dover,
New York, 1979).
\bibitem{deo1} P. Singha Deo, Phys. Rev. B {\bf 53}, 15447 (1996);
P. A. Sreeram and P. Singha Deo, Physica B {\bf 228}, 345 (1996).
\bibitem{the} P.Singha Deo and A.M.Jayannavar, Mod. Phys. Lett. B {\bf 10},
787 (1996);
C.M.Ryu et al, Phys. Rev. B {\bf 58}, 3572 (1998);
Hongki Xu et al, Phys. Rev. B, {\bf 57}, 11903 (1998);
\bibitem{deo2} P.Singha Deo, Solid St. Communication {\bf 107}, 69 (1998);
\bibitem{lee} H.-W.Lee, Phys. Rev. Lett., {\bf 82}, 2358 (1999).
\bibitem{tan} T. Taniguchi and M. B\"uttiker, Phys. Rev. B {\bf 60},
13814 (1999), and references therein.
\bibitem{sch} R. Schuster et al, Nature {\bf 385}, 417 (1997).
\bibitem{bay} B. F. Bayman and C. J. Mehoke, Am. Journ. of Phys.
{\bf 51} 875(1983); W. Porod, Z. Shao and C. S. Lent, Phys. Rev. B
{\bf 48}, 8495(1993) and references therein.
\bibitem{jay1} A. M. Jayannavar and
T.P.Pareek, Phys. Rev. B {\bf 54}, 6376 (1996) and references
therein.
\bibitem{jay2} A. M. Jayannavar, private communication (1994).
\bibitem{gan} A. Gangopadhyaya, A. Pagnamenta and U. Sukhatme,
Journ. of Phys. A {\bf 28}, 5331 (1995); M. B{\"u}ttiker, Y. Imry
and M. Ya Azbel, Phys. Rev. A {\bf 30}, 1982 (1984).
\end{thebibliography}
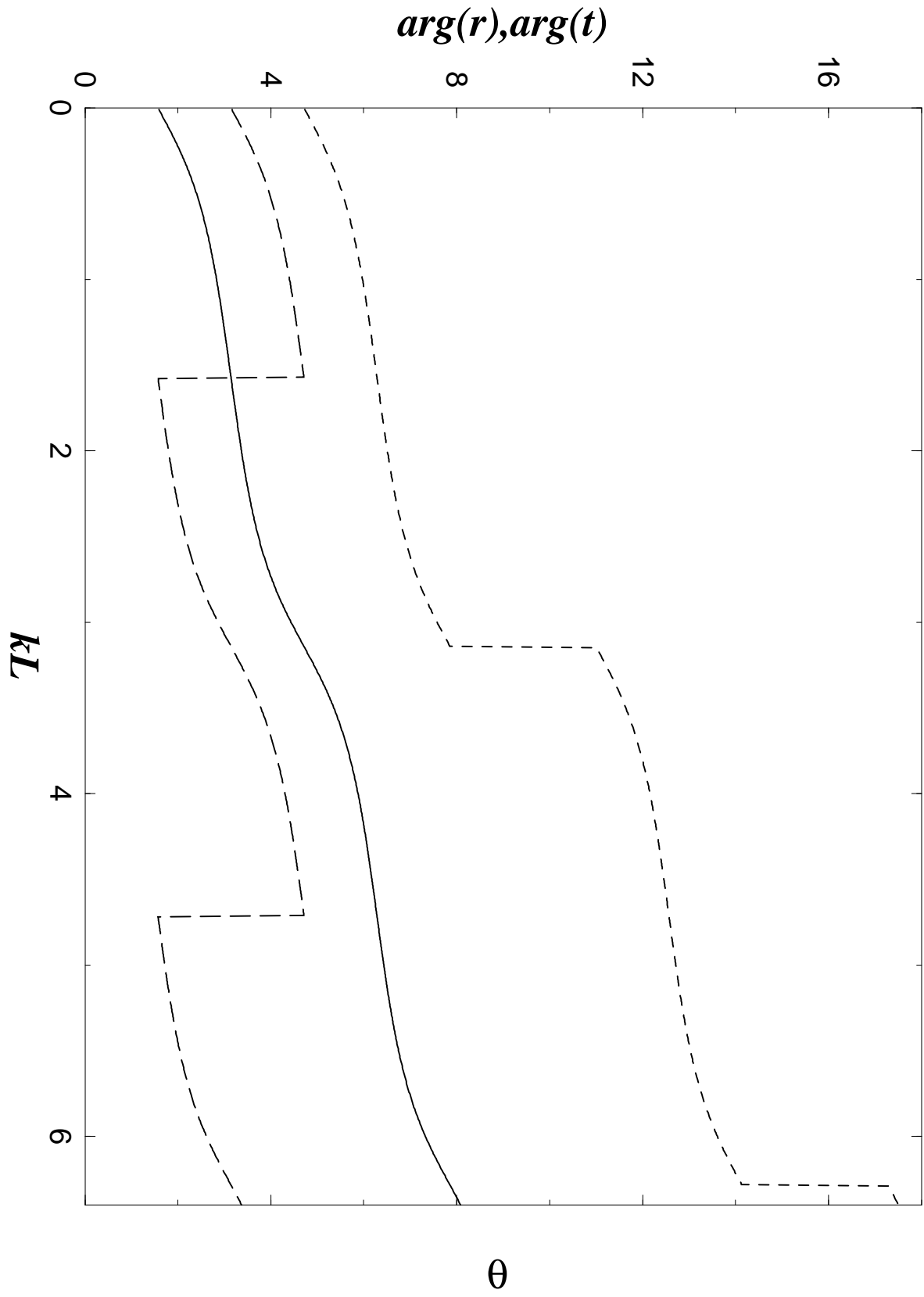
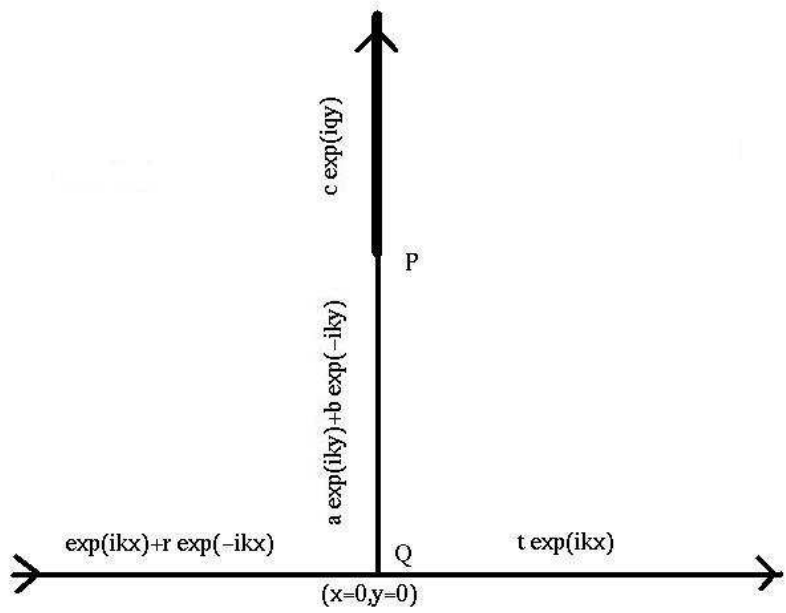
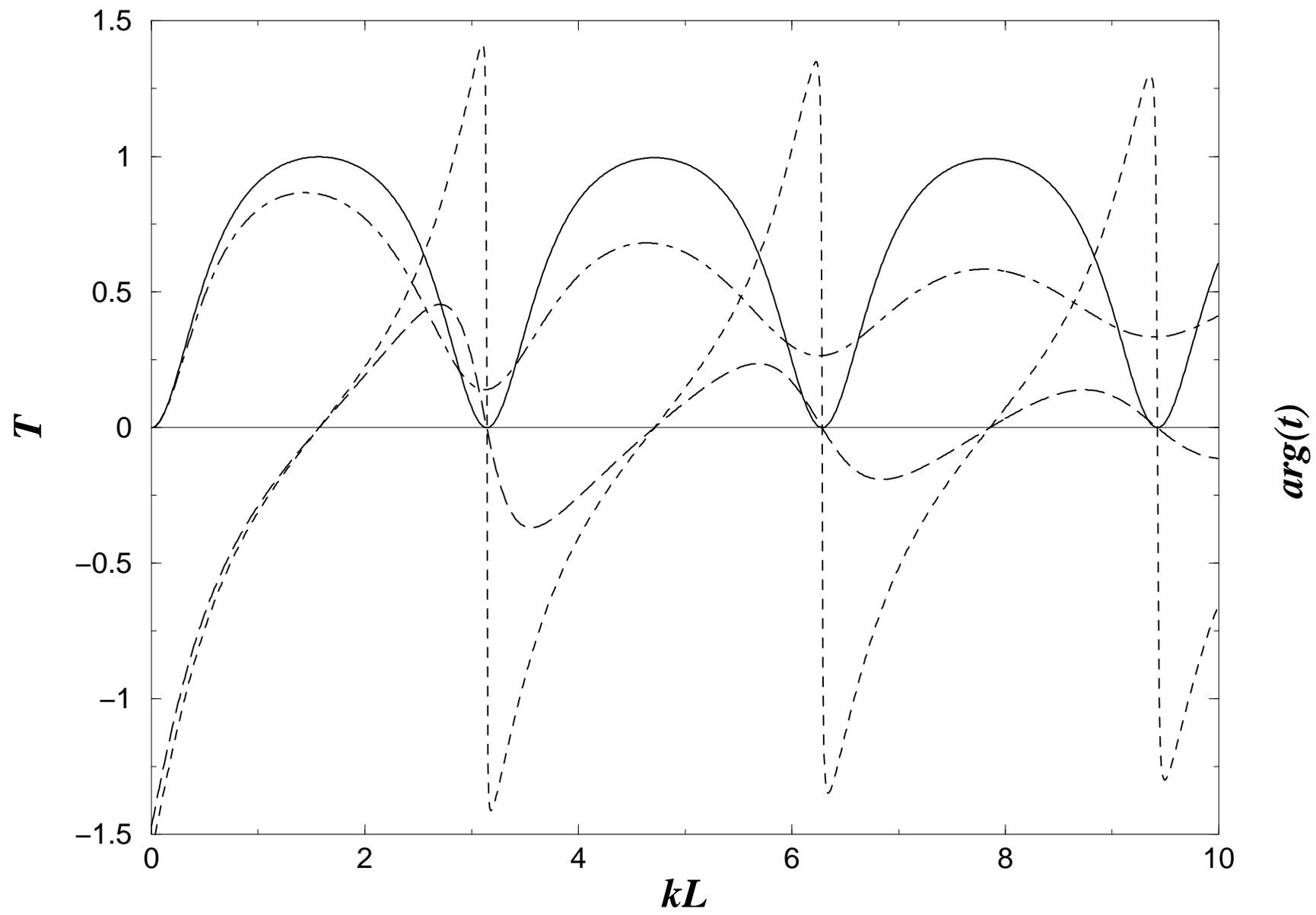
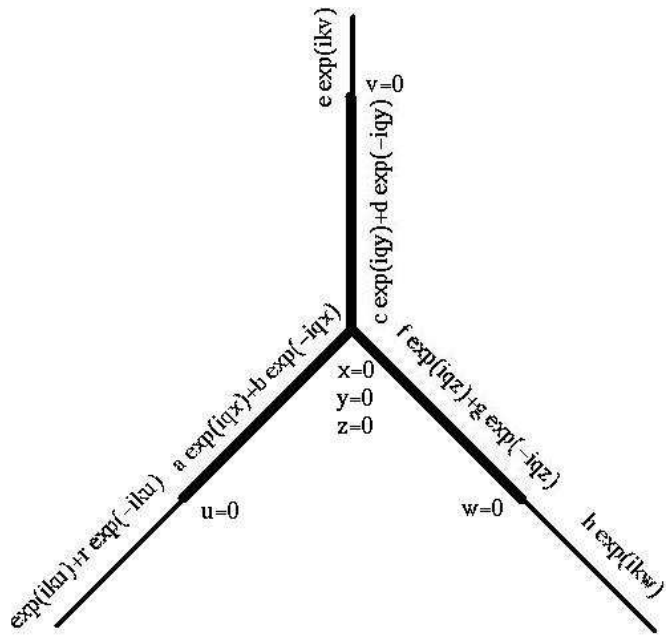
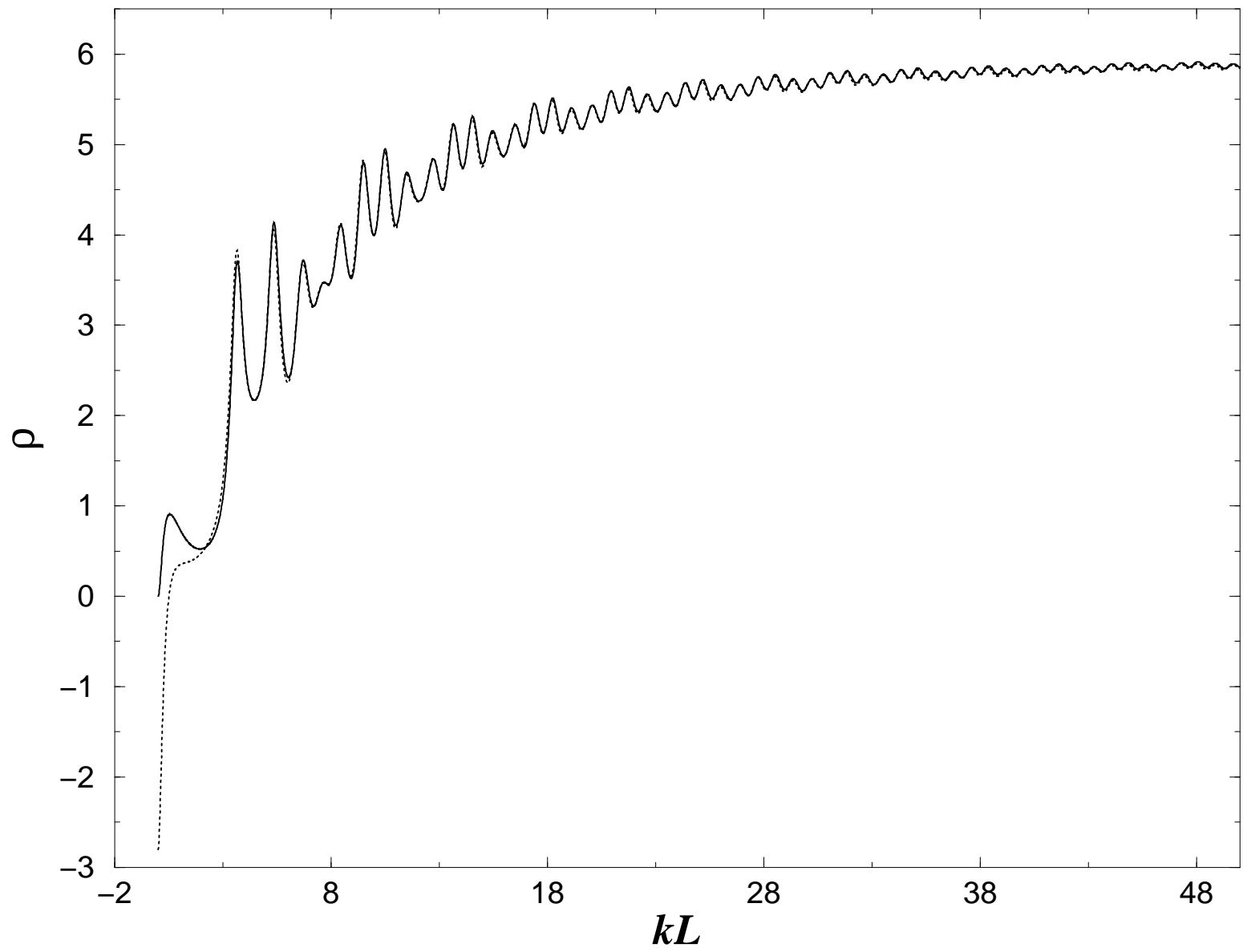
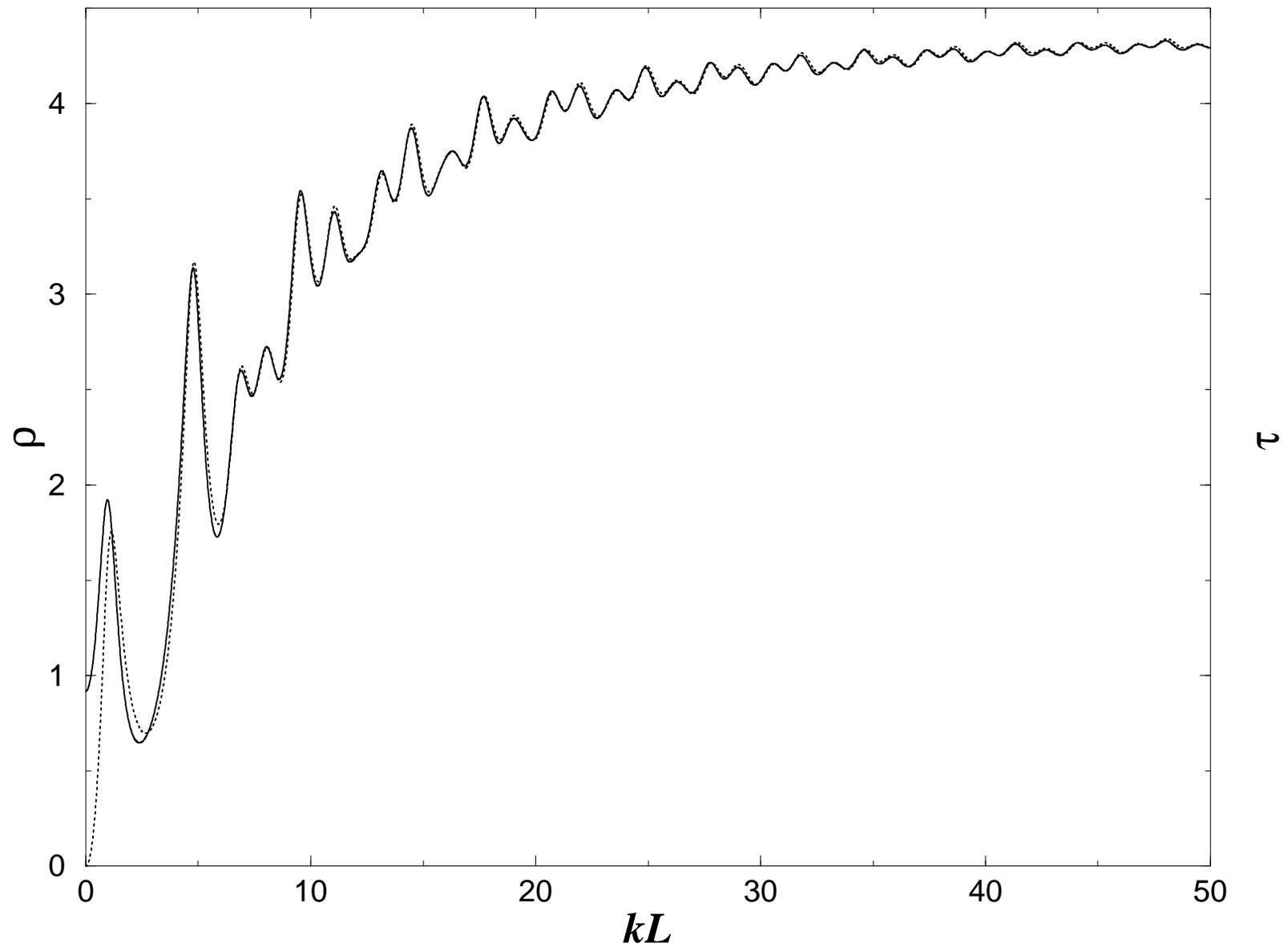
\end{document}